\pdfoutput=1
\documentclass[prl,letterpaper,twocolumn,10pt,aps,notitlepage,superscriptaddress]{revtex4-1}

\usepackage{amsfonts}
\usepackage{bm,graphicx}
\usepackage{hyperref}

\hypersetup{colorlinks, linkcolor = [rgb]{0,0.0,0.75}, citecolor = [rgb]{0,0.0,0.75}, urlcolor = [rgb]{0,0.0,0.75}}

\newcommand{\nb}{\phantom{0}}
\newcommand{\wm}{\phantom{-}}

\begin{document}

\title{Calculation of $\bm{B^0\to K^{*0} \mu^+\mu^-}$ and $\bm{B_s^0 \to \phi\, \mu^+\mu^-}$ observables \\ using form factors from lattice QCD}

\author{Ronald R.~Horgan}
\affiliation{\mbox{Department of Applied Mathematics and Theoretical Physics, University of Cambridge, Cambridge CB3 0WA, UK}}
\author{Zhaofeng Liu}
\affiliation{Institute of High Energy Physics and Theoretical Physics Center for Science Facilities, Chinese Academy of Sciences, Beijing 100049, China}
\author{Stefan Meinel}
\email{smeinel@mit.edu}
\affiliation{Center for Theoretical Physics, Massachusetts Institute of Technology, Cambridge, MA 02139, USA}
\author{Matthew Wingate}
\affiliation{\mbox{Department of Applied Mathematics and Theoretical Physics, University of Cambridge, Cambridge CB3 0WA, UK}}

\date{April 17, 2014}

\begin{abstract}
We calculate the differential branching fractions and angular distributions of the rare decays $B^0 \to K^{*0} \mu^+\mu^-$
and $B_s^0 \to \phi\, \mu^+\mu^-$, using for the first time form factors from unquenched lattice QCD. We focus on the
kinematic region where the $K^*$ or $\phi$ recoils
softly; there the newly available form factors are most precise and the nonlocal matrix elements can be included via an operator
product expansion. Our results for the differential branching fractions calculated in the Standard Model are higher than the
experimental data. We consider the possibility that the deviations are caused by new physics, and perform a fit of the
Wilson coefficients $C_9$ and $C_9^\prime$ to the experimental data for multiple $B^0 \to K^{*0} \mu^+\mu^-$ and
$B_s^0 \to \phi\, \mu^+\mu^-$ observables. In agreement with recent results from complementary studies,
we obtain $C_9-C_9^{\rm SM}=-1.0\pm 0.6$ and $C_9^\prime = 1.2 \pm 1.0$, whose deviations
from zero would indicate the presence of non-standard fundamental interactions.
\end{abstract}

\maketitle

Decays involving the transition of a bottom quark to a strange quark are highly suppressed in the Standard Model. Contributions
from non-standard interactions could therefore be significant, causing observable changes in the decay rates and angular distributions.
The search for such discrepancies is one of the most important routes to discovering what might lie beyond our
current model of fundamental particle physics, and complements efforts to directly produce non-standard particles.
Because of quark confinement, the $b\to s$ transitions are being observed with hadronic initial and final states.
Among the cases that have been measured experimentally \cite{Amhis:2012bh}, the decay $B \to K^*
\ell^+ \ell^-$, (where $\ell$ is an electron or muon) is proving to be particularly
powerful in looking for physics beyond the Standard Model \cite{Altmannshofer:2008dz,
Alok:2009tz, Bobeth:2010wg, Alok:2010zd, Alok:2011gv, Bobeth:2011gi, Becirevic:2011bp, Altmannshofer:2011gn, Matias:2012xw,
Beaujean:2012uj, Altmannshofer:2012az, Bobeth:2012vn}.

The LHCb Collaboration recently
published new precision measurements of the decay $B \to K^*\mu^+ \mu^-$, and one of the observables shows a significant deviation
from the Standard Model predictions \cite{Aaij:2013qta}.
There is currently an intense effort to understand this discrepancy, which could be a manifestation of new physics
\cite{Descotes-Genon:2013wba, Altmannshofer:2013foa, Gauld:2013qba, Hambrock:2013zya, Buras:2013qja, Gauld:2013qja, Datta:2013kja, Beaujean:2013soa}.
Previous calculations of the matrix elements that relate the underlying $b\to s$ interactions and the
hadronic observables are reliable only in the kinematic region of high recoil (large $K^*$ momentum
in the $B$ rest frame), and consequently it was in this region that a discrepancy was found.
In the low-recoil region, numerical lattice QCD computations must be performed.
We recently completed the first unquenched lattice QCD calculation of the form factors
that parametrize the hadronic matrix elements
relevant for $B \to K^*\ell^+ \ell^-$ and $B_s \to \phi\,\ell^+ \ell^-$ \cite{Horgan:2013hoa}. In this Letter,
we investigate the consequences of using these results in combination
with experimental data. We find that hints of deviations from the Standard Model
are present also in the low-recoil region, and a better fit of the data is obtained
by allowing non-standard interactions \mbox{consistent} with those suggested to explain the
aforementioned anomaly at high recoil.

At hadronic energy scales, $b\to s\gamma$ and $b\to s \ell^+\ell^-$  transitions can be described using an effective Hamiltonian
of the form \cite{Grinstein:1988me,Grinstein:1990tj,Misiak:1992bc,Buras:1993xp,Buras:1994dj,Buchalla:1995vs,Chetyrkin:1996vx,Bobeth:1999mk}
\begin{equation}
\mathcal{H}_{\rm eff} = -\frac{4 G_F}{\sqrt{2}}V_{tb}V_{ts}^* \: \sum_{i} \left[ \: C_i O_i + C_i^{\prime} O_i^{\prime} \: \right], \label{eq:Heff}
\end{equation}
where $O_i^{(\prime)}$ are local operators and $C_i^{(\prime)}$ are the corresponding Wilson coefficients,
encoding the physics at the electroweak energy scale and beyond.
The operators
\begin{eqnarray}
{O_7^{(\prime)}}    &=& e\: m_b /(16\pi^2)\:\: \bar{s} \sigma_{\mu\nu} {P_{R(L)}} b \:\:\: F^{\mu\nu}, \label{eq:O7}  \\
{O_9^{(\prime)}}    &=& e^2/(16\pi^2)\:\: \bar{s} \gamma_\mu {P_{L(R)}} b\:\:\: \bar{\ell} \gamma^\mu \ell, \label{eq:O9}  \\
{O_{10}^{(\prime)}} &=& e^2/(16\pi^2)\:\: \bar{s} \gamma_\mu {P_{L(R)}} b\:\:\: \bar{\ell} \gamma^\mu \gamma_5 \ell, \label{eq:O10}
\end{eqnarray}
where $F^{\mu\nu}$ is the electromagnetic field strength tensor,
give the leading contributions to the decays we will discuss in this work. The operators $O_{1...6}^{(\prime)}$ are
four-quark operators, and $O_8^{(\prime)}$ contains the gluon field strength tensor. The primed operators differ from
the unprimed operators in their chirality [$P_{R,L}=(1\pm\gamma_5)/2$]; the Standard Model predicts that their Wilson
coefficients, $C_i^{\prime}$, are negligibly small.

The utility of the decay $B \to K^* (\to K\,\pi) \ell^+ \ell^-$ is that all 
six Dirac structures in Eqs.~(\ref{eq:O7}-\ref{eq:O10}) have nonzero matrix elements, and the 
angular distribution can be used to disentangle them. In the narrow-width approximation \cite{Kruger:1999xa, Kim:2000dq},
the kinematics of the quasi-four-body decay $\bar{B}^0\to \bar{K}^{*0}(\to K^-\pi^+) \ell^+\ell^-$ is described by four variables:
the invariant mass of the lepton pair, $q^2$, and the three angles $\theta_\ell$, $\theta_{K^*}$, $\phi$, defined here
as in Ref.~\cite{Altmannshofer:2008dz}.
In this approximation, the general form of the decay distribution is \cite{Kruger:1999xa, Kim:2000dq, Faessler:2002ut, Kruger:2005ep, Altmannshofer:2008dz}
\begin{eqnarray}
\nonumber && \frac{\mathrm{d}^4 \Gamma}{\mathrm{d} q^2\: \mathrm{d}\cos\theta_\ell \: \mathrm{d}\cos\theta_{K^*}\: \mathrm{d}\phi} \\
\nonumber && = \frac{9}{32\pi} \Big[ \phantom{+} I_1^s \sin^2\theta_{K^*} + I_1^c \cos^2\theta_{K^*} \\
\nonumber && \hspace{8ex} + \: (I_2^s \sin^2\theta_{K^*} + I_2^c \cos^2\theta_{K^*}) \cos 2\theta_\ell  \\
\nonumber && \hspace{8ex} + \: I_3 \sin^2\theta_{K^*} \sin^2\theta_\ell \cos 2\phi \\
\nonumber && \hspace{8ex} + \: I_4 \sin 2\theta_{K^*} \sin 2\theta_\ell \cos\phi \\
\nonumber && \hspace{8ex} + \: I_5 \sin 2\theta_{K^*} \sin\theta_\ell \cos\phi \\
\nonumber && \hspace{8ex} + \: (I_6^s \sin^2\theta_{K^*} + I_6^c \cos^2\theta_{K^*} ) \cos\theta_\ell \\
\nonumber && \hspace{8ex} + \: I_7 \sin 2\theta_{K^*} \sin\theta_\ell \sin\phi \\
\nonumber && \hspace{8ex} + \: I_8 \sin 2\theta_{K^*} \sin 2\theta_\ell \sin\phi \\
          && \hspace{8ex} + \: I_9 \sin^2\theta_{K^*} \sin^2\theta_\ell \sin 2\phi \Big], \label{eq:angulardist}
\end{eqnarray}
where the coefficients $I_i^{(a)}$ depend only on $q^2$.
Integrating over the angles, one obtains the differential decay rate $\mathrm{d}\Gamma/\mathrm{d} q^2 = \frac34 (2 I_1^s+I_1^c)-\frac14 (2 I_2^s+ I_2^c)$.
The angular distribution of the CP-conjugated mode $B^0\to K^{*0}(\to K^+\pi^-)\ell^+\ell^-$ is obtained from
Eq.~(\ref{eq:angulardist}) through the replacements $I_{1,2,3,4,7}^{(a)} \to \bar{I}_{1,2,3,4,7}^{(a)}$,
$I_{5,6,8,9}^{(a)} \to -\bar{I}_{5,6,8,9}^{(a)}$ \cite{Altmannshofer:2008dz}. Normalized CP averages and CP asymmetries of the angular coefficients
are then defined as follows \cite{Altmannshofer:2008dz}:
\begin{equation}
 S_i^{(a)}=\frac{I_i^{(a)}+\bar{I}_i^{(a)}}{\mathrm{d}(\Gamma+\bar{\Gamma})/\mathrm{d} q^2}, \hspace{3ex}
 A_i^{(a)}=\frac{I_i^{(a)}-\bar{I}_i^{(a)}}{\mathrm{d}(\Gamma+\bar{\Gamma})/\mathrm{d} q^2}. \label{eq:CPasym}
\end{equation}
The experiments actually yield results for binned observables $\langle S_i^{(a)} \rangle$ and $\langle A_i^{(a)} \rangle$,
given by the ratios of $q^2$-integrals of numerator and denominator in Eq.~(\ref{eq:CPasym}).

The observables $\langle S_{4,5,7,8} \rangle$ and the ratios
\begin{equation}
 \langle P_{4,5,6,8}^\prime \rangle  = \frac{\langle S_{4,5,7,8} \rangle }{2\sqrt{-\langle S_2^c \rangle \langle S_2^s \rangle}} \label{eq:Piprime}
\end{equation}
(note the different indices on the left-hand and right-hand sides) have recently been measured for the first time by
the LHCb Collaboration in the decay $\bar{B}^0 \to \bar{K}^{*0}(\to K^-\pi^+)\mu^+\mu^-$ (and its CP-conjugate) \cite{Aaij:2013qta}.
The ratios (\ref{eq:Piprime}) are designed to reduce hadronic uncertainties at low $q^2$ \cite{DescotesGenon:2012zf}.
For $P_5^\prime$, a significant discrepancy between the LHCb result and the Standard-Model prediction of
Ref.~\cite{Descotes-Genon:2013vna} was found in the bin $4.30\:{\rm GeV^2}\leq q^2 \leq 8.68\:{\rm GeV}^2$ \cite{Aaij:2013qta}.
In Ref.~\cite{Descotes-Genon:2013wba} it was suggested that this discrepancy, as well as some smaller deviations
in other observables, can be explained by a negative new-physics contribution to the Wilson coefficient $C_9$; specifically,
$C_9=C_9^{\rm SM}+C_9^{\rm NP}$, where $C_9^{\rm SM}\approx 4$ and $C_9^{\rm NP}\approx -1.5$. The authors of
Ref.~\cite{Altmannshofer:2013foa} performed global fits of the latest experimental data in multiple $b\to s$ decay channels,
allowing various subsets of the Wilson coefficients to deviate from their respective Standard-Model values.
Allowing two Wilson coefficients to deviate, the biggest reduction in $\chi^2$ was obtained for
\begin{equation}
 C_9^{\rm NP} = -1.0\pm0.3, \hspace{2ex}C_9^\prime = 1.0 \pm 0.5. \label{eq:AltmannshoferGF}
\end{equation}
Such large effects in $C_9$ and $C_9^\prime$ can arise in models with flavor-changing neutral gauge
bosons ($Z^\prime$) in the few-TeV mass range \cite{Descotes-Genon:2013wba, Altmannshofer:2013foa, Gauld:2013qba, Buras:2013qja, Gauld:2013qja},
and in models that generate new four-quark operators of scalar and pseudoscalar type \cite{Datta:2013kja}.

Extractions of Wilson coefficients from the experimental data require knowledge of the matrix elements of the
operators $O_i^{(\prime)}$ in nonperturbative QCD. The analyses discussed above are
based on calculations of the $B \to K^*$ matrix elements using light-cone sum rules \cite{Colangelo:2000dp,Ball:2004rg,Khodjamirian:2010vf}
and QCD factorization \cite{Beneke:2001at}. These calculations are limited to the low-$q^2$ (high recoil) region.
On the other hand, the experiments cover the entire kinematic range
$4 m_\ell^2 < q^2 < (m_B-m_{K^*})^2 \approx 19\:{\rm GeV^2}$, and changes in $C_9^{(\prime)}$ will also affect the
high-$q^2$ region. We have recently completed the first lattice QCD calculation of the complete set
of form factors giving the $B \to K^*$ and $B_s \to \phi$ matrix elements of the operators $O_7^{(\prime)}$, $O_9^{(\prime)}$,
and $O_{10}^{(\prime)}$ in the high-$q^2$ region \cite{Horgan:2013hoa}. In the following, we use these results to calculate the differential branching
fractions and the angular observables for the decays $\bar{B}^0\to \bar{K}^{*0}(\to K^-\pi^+) \mu^+\mu^-$
and $\bar{B}_s^0\to \phi(\to K^- K^+)\mu^+\mu^-$.

In the narrow-width approximation, the $\bar{B}^0\to K^-\pi^+ \mu^+\mu^-$ decay amplitude can be written in terms of the
$\bar{B}^0\to \bar{K}^{*0} \mu^+\mu^-$ decay amplitude as explained in Ref.~\cite{Kruger:1999xa}.
This amplitude takes the form
\begin{equation}
 \mathcal{M} = \frac{G_F\, \alpha}{\sqrt{2}\pi}V_{tb}V_{ts}^*
 \Big[  (\mathcal{A}_\mu + \mathcal{T}_\mu) \bar{u}_\ell \gamma^\mu v_\ell
 + \mathcal{B}_\mu  \bar{u}_\ell\gamma^\mu\gamma_5 v_\ell \Big], \label{eq:amplitude}
\end{equation}
with the local hadronic matrix elements
\begin{eqnarray}
\nonumber \mathcal{A}_\mu &=& - \frac{2m_b}{q^2}\: q^\nu \langle \bar{K}^* | \: \bar{s}\, i\sigma_{\mu\nu}  ({C_7} P_R + {C_7^\prime} P_L) b \: | \bar{B} \rangle \\
                          & & +\: \langle \bar{K}^* |\: \bar{s} \gamma_\mu ({C_9} P_L + {C_9^\prime}P_R) b\:      | \bar{B} \rangle, \label{eq:LM1}\\
          \mathcal{B}_\mu &=&     \langle \bar{K}^* |\: \bar{s} \gamma_\mu ({C_{10}}P_L + {C_{10}^\prime}P_R) b\: | \bar{B} \rangle, \label{eq:LM2}
\end{eqnarray}
and the nonlocal hadronic matrix element
\begin{equation}
\mathcal{T}_\mu = \frac{-16 i \pi^2}{q^2} \!\sum_{i=1...6; 8}\!  C_i\int \! \mathrm{d}^4 x \:\:
e^{iq\cdot x}\,\langle \bar{K}^* | \: \mathsf{T} \:  O_i(0) \: j_\mu(x) \: | \bar{B} \rangle. \label{eq:NLME}
\end{equation}
In Eq.~(\ref{eq:NLME}), $j_\mu(x)$ denotes the quark electromagnetic current. Near $q^2 = m_{J/\psi(1S)}^2, m_{\psi(2S)}^2$,
the contributions from $O_1$ and $O_2$ in $\mathcal{T}_\mu$ are resonantly enhanced, preventing reliable theoretical
calculations in these regions. At high $q^2$ ($\sim m_b^2$), $\mathcal{T}_\mu$ can be expanded in an operator product
expansion (OPE), with the result \cite{Grinstein:2004vb}
\begin{eqnarray}
\nonumber \mathcal{T}_\mu &=& - T_7(q^2) \frac{2m_b}{q^2}\: q^\nu\langle \bar{K}^* | \, \bar{s}\, i\sigma_{\mu\nu}  P_R  b \, | \bar{B} \rangle \\
\nonumber                 & & + T_9(q^2)  \langle \bar{K}^* | \bar{s} \gamma_\mu P_L  b  | \bar{B} \rangle
                              + \frac{1}{2q^2} \sum_{i=1}^5  B_i   \langle \bar{K}^* |  O_{i\mu}^{(-1)}   | \bar{B} \rangle \\
                          & & + \:\mathcal{O}(\Lambda^2/m_b^2,\:\: m_c^4/q^4 ). \label{eq:NLMEOPE}
\end{eqnarray}
(See also Ref.~\cite{Beylich:2011aq} for an alternative version of the OPE.) In Eq.~(\ref{eq:NLMEOPE}),
the $O_{i\mu}^{(-1)}$ are dimension-4 operators \mbox{containing} a derivative, and $T_{7,9}(q^2)=C_{7,9}^{\rm eff}(q^2)-C_{7,9}$
with $C_{7,9}^{\rm eff}(q^2)$ given by Eqs.~(3.9) and (3.10) of Ref.~\cite{Bobeth:2010wg}.

The matrix elements $\langle \bar{K}^* |\, \bar{s} \Gamma b\, | \bar{B} \rangle$ (and analogously for $\bar{B}_s \to \phi$)
in Eqs.~(\ref{eq:LM1}), (\ref{eq:LM2}), and (\ref{eq:NLMEOPE}) can be written in terms of the seven form factors $V$,
$A_0$, $A_1$, $A_{12}$, $T_1$, $T_2$, and $T_{23}$ \cite{Horgan:2013hoa}. We describe the dependence of the form factors on $q^2$
using the simplified series expansion \cite{Bourrely:2008za}. The corresponding parameters were obtained by fitting the
lattice QCD data, and are given in Tables VII - XI of Ref.~\cite{Horgan:2013hoa}. The matrix elements of the dimension-4 operators
in Eq.~(\ref{eq:NLMEOPE}) have not yet been calculated in lattice QCD, and we will neglect this term. This introduces a
small systematic uncertainty of order $\alpha_s \Lambda/m_b \sim 2\%$ \cite{Grinstein:2004vb}.

We take the Standard-Model values of the Wilson coefficients $C_{1,2,...,10}$, calculated at next-to-next-to-leading-logarithmic
order, from Ref.~\cite{Altmannshofer:2008dz}. Following the same reference, we set $\alpha_s(m_b)=0.214$,
$m_c(m_c)=1.3\:{\rm GeV}$, and $m_b(m_b)=4.2\:{\rm GeV}$. We evaluate the electromagnetic coupling at $\mu=m_b$, corresponding to
$\alpha = 1/133$, which minimizes higher-order electroweak corrections \cite{Bobeth:2003at}. We take the hadron masses from
the Particle Data Group \cite{Beringer:1900zz} and use the mean life times $\tau_{B^0}=1.519(7)\:{\rm ps}$ and
$\tau_{B_s^0}=1.516(11)\:{\rm ps}$ from Ref.~\cite{Amhis:2012bh}. We take $|V_{tb} V_{ts}^*|=0.04088(57)$ from the
Summer 2013 Standard-Model fit of Ref.~\cite{UTFit}.

While the decay $\bar{B}^0\to \bar{K}^{*0}(\to K^-\pi^+) \mu^+\mu^-$ is self-tagging, the final state of
$\bar{B}_s^0\to \phi(\to K^- K^+)\ell^+\ell^-$ does not determine whether it resulted from the decay of a $\bar{B}_s^0$
or a $B_s^0$ meson. Therefore, we calculate the time-integrated untagged average over the $\bar{B}_s^0$ and $B_s^0$
decay distributions, including the effects of $\bar{B}_s^0$-$B_s^0$ mixing as explained in Ref.~\cite{Bobeth:2008ij}.
We use the width difference $\Delta \Gamma_s=0.081(11)\:{\rm ps}^{-1}$ \cite{Amhis:2012bh}.

\begin{figure*}
\hspace{1cm}\includegraphics[width=\linewidth]{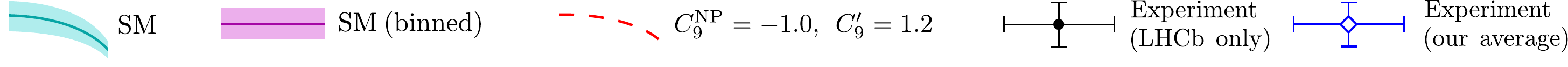}

\includegraphics[width=0.24\linewidth]{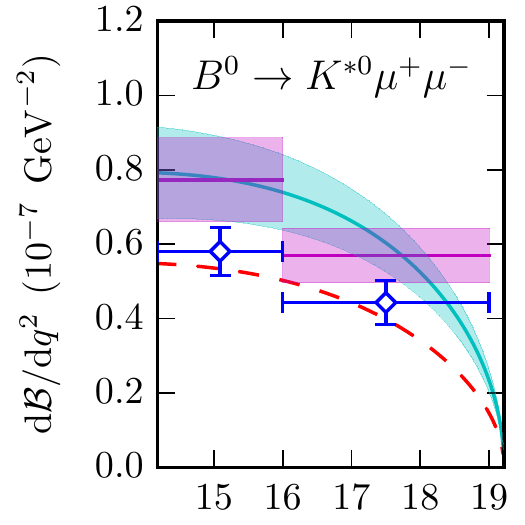} \hfill
\includegraphics[width=0.24\linewidth]{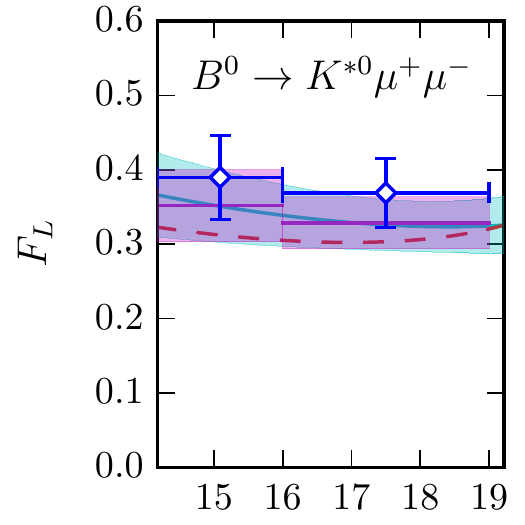} \hfill 
\includegraphics[width=0.24\linewidth]{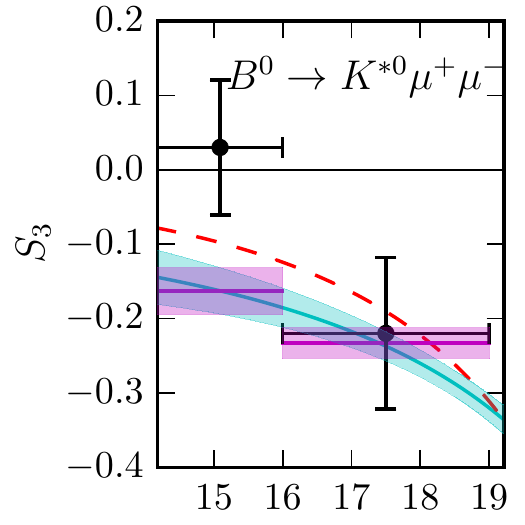} \hfill
\includegraphics[width=0.24\linewidth]{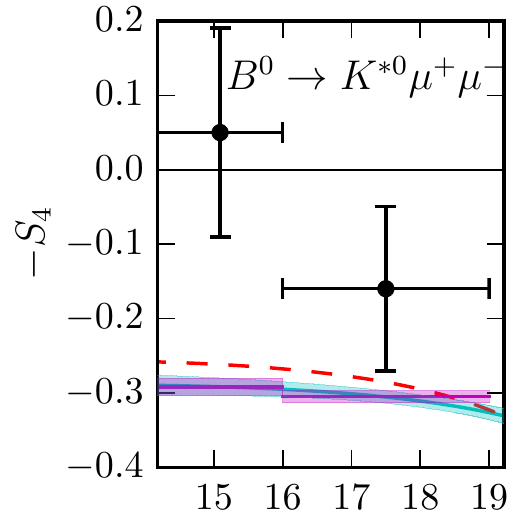}

\includegraphics[width=0.24\linewidth]{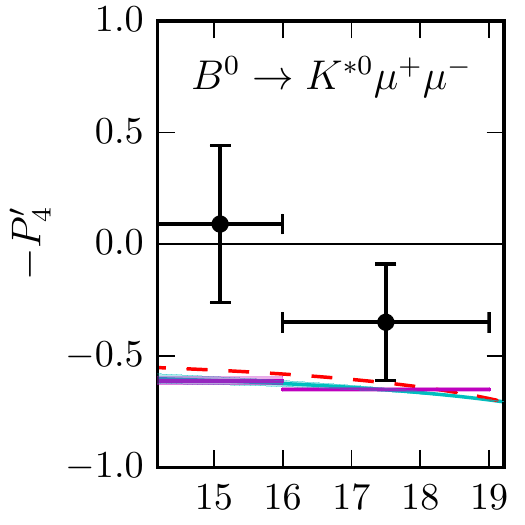} \hfill 
\includegraphics[width=0.24\linewidth]{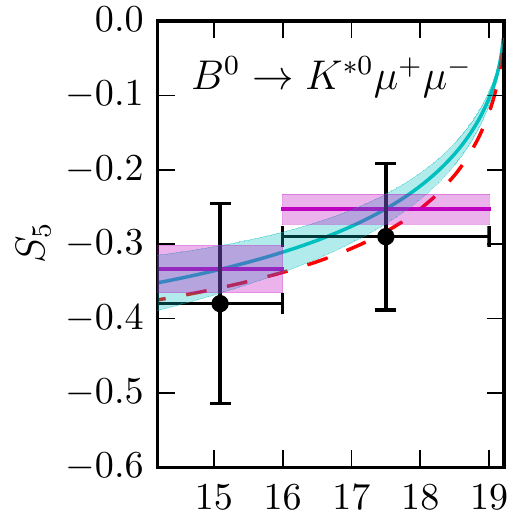} \hfill
\includegraphics[width=0.24\linewidth]{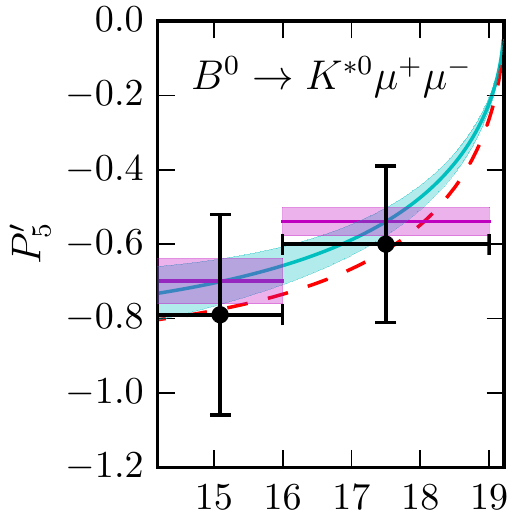} \hfill
\includegraphics[width=0.24\linewidth]{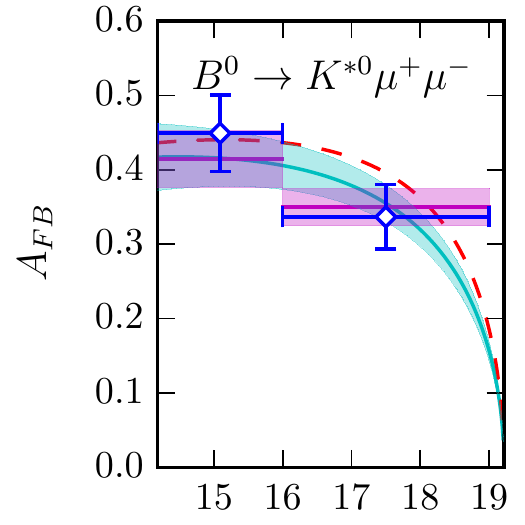}

\includegraphics[width=0.24\linewidth]{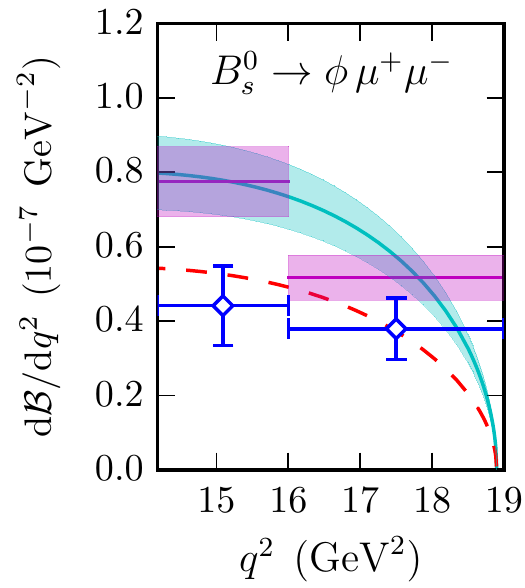} \hfill
\includegraphics[width=0.24\linewidth]{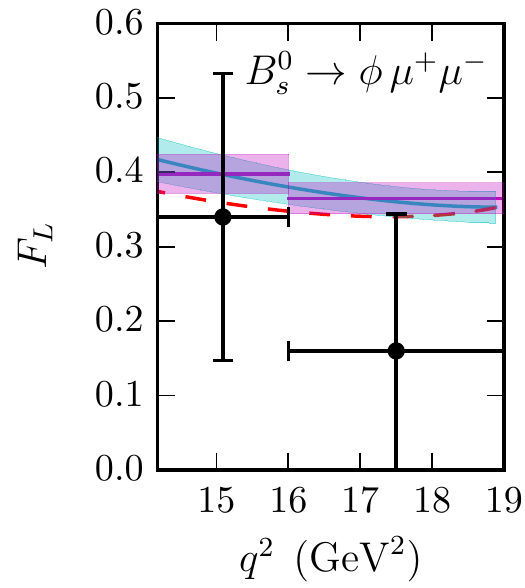} \hfill
\includegraphics[width=0.24\linewidth]{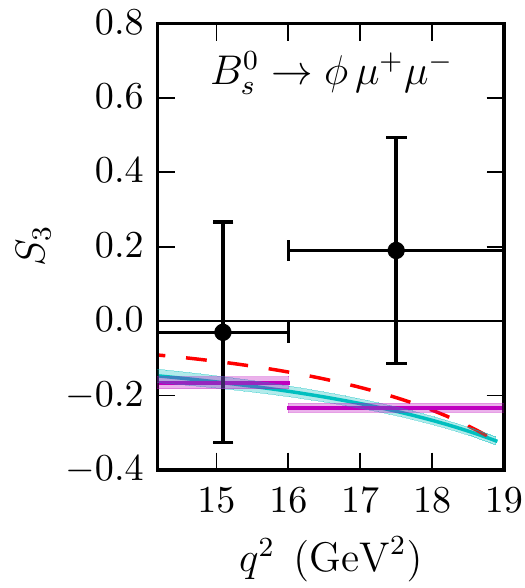} \hfill
\includegraphics[width=0.24\linewidth]{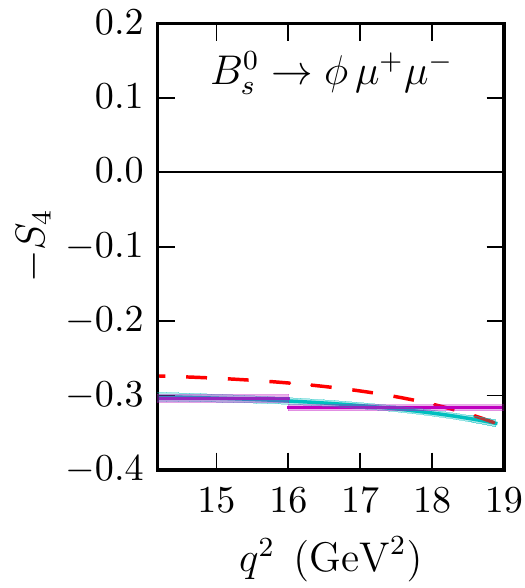}

\caption{\label{fig:observables}Observables for the decays $B^0\to K^{*0}\mu^+\mu^-$ (upper two rows) and
$B_s^0\to \phi\,\mu^+\mu^-$ (bottom row; untagged averages over the $\bar{B}_s^0$ and $B_s^0$ distributions). The solid curves show
our theoretical results in the Standard Model; the shaded areas give the corresponding total uncertainties (with and without binning).
The dashed curves correspond to the new-physics fit result $C_9=C_9^{{\rm SM}}-1.0$, $C_9^\prime=1.2$ (the uncertainties of the
dashed curves are not shown for clarity). We also show our averages of results from the CDF, LHCb, CMS, and ATLAS experiments
\cite{CDF2012, Aaij:2013iag, Aaij:2013aln, Aaij:2013qta, ATLAS2013} (note that $S_4^{(\rm LHCb)}=-S_4$ and
$P_4^{\prime(\rm LHCb)}=-P_4^\prime$).}
\end{figure*}

Our results for the differential branching fractions $\mathrm{d}\mathcal{B}/\mathrm{d}q^2=\tau_{B^0_{(s)}}\mathrm{d}\Gamma/\mathrm{d}q^2$
and the angular observables $F_L$, $S_3$, $S_4$, $P_4^\prime$, $S_5$, $P_5^\prime$, $A_{FB}$, where $F_L=-S_2^c$ and
$A_{FB}=(-3/8)(2S_6^s+S_6^c)$, are shown in Fig.~\ref{fig:observables} (the observables $S_{7,8,9}$ as well as the CP
asymmetries $A_i^{(a)}$ are expected to be close to zero in the Standard Model). The shaded bands in Fig.~\ref{fig:observables}
indicate the total theoretical uncertainty, originating from the following sources: the statistical/fitting and systematic
uncertainty in the form factors \cite{Horgan:2013hoa}, an estimated 2\% uncertainty in the values of the Wilson coefficients
$C_i$ \cite{Altmannshofer2percent}, the uncertainties in the $B^0$ and $B_s^0$ meson mean life times, the uncertainty
in $|V_{tb}V_{ts}^*|$, and an estimated additional 5\% systematic uncertainty in the vector amplitude
$(\mathcal{A}_\mu+\mathcal{T}_\mu)$ in Eq.~(\ref{eq:amplitude}), which is introduced by the truncation of the OPE and
duality violations \cite{Grinstein:2004vb,Beylich:2011aq}. Note that $S$-wave pollution is expected to be negligible
at large $q^2$ \cite{Becirevic:2012dp}.

In Fig.~\ref{fig:observables}, we also show experimental results, which are given for the bins
$14.18\:{\rm GeV}^2 < q^2 < 16\:{\rm GeV}^2$ (bin 1) and $16\:{\rm GeV}^2 < q^2 < 19\:{\rm GeV}^2$ (bin 2).
Some of the observables have only been measured by LHCb \cite{Aaij:2013iag, Aaij:2013aln, Aaij:2013qta}.
For the $B_s^0 \to \phi \mu^+ \mu^-$ branching fraction, we averaged the results from LHCb \cite{Aaij:2013aln}
and CDF \cite{CDF2012}. For the $B^0 \to K^{*0}\mu^+\mu^-$ branching fraction, we averaged the results from LHCb \cite{Aaij:2013iag},
CMS \cite{Chatrchyan:2013cda}, and CDF (bin 1 only, due to different upper $q^2$ limit in bin 2) \cite{CDF2012}.
For $A_{FB}$ and $F_L$, we additionally included the ATLAS results \cite{ATLAS2013} in the average.
Our binned theoretical results are given in Table \ref{tab:binnedresults} and are also shown in Fig.~\ref{fig:observables}.

\begin{table}

\begin{tabular}{ccccc}
\hline\hline
Observable                 & \hspace{2ex} &  $[14.18,\:16.00]$   & \hspace{2ex} & $[16.00,\:19.00]$ \\
\hline
$B^0 \to K^{*0} \mu^+\mu^-$               &&                     &&                                 \\
$\langle\mathrm{d}\mathcal{B}/\mathrm{d}q^2 \rangle \:\:(10^{-7}\:{\rm GeV^{-2}})$
                                          && $\wm0.77(11)\nb$    && $\wm0.569(74)\nb$               \\
$\langle F_L \rangle$                     && $\wm0.352(49)$      && $\wm0.329(35)\nb$               \\
$\langle S_3 \rangle$                     && $-0.163(31)$        && $-0.233(20)\nb$                 \\
$\langle S_4 \rangle$                     && $\wm0.292(12)$      && $\wm0.3051(84)$               \\
$\langle P_4^\prime \rangle$              && $\wm0.613(18)$      && $\wm0.6506(84)$                 \\
$\langle S_5 \rangle$                     && $-0.333(32)$        && $-0.253(20)\nb$                 \\
$\langle P_5^\prime \rangle$              && $-0.700(61)$        && $-0.539(38)\nb$                 \\
$\langle A_{FB} \rangle$                  && $\wm0.414(38)$      && $\wm0.350(25)\nb$               \\
\hline
$B_s^0 \to \phi\, \mu^+\mu^-$             &&                     &&                                 \\
$\langle \mathrm{d}\mathcal{B}/\mathrm{d}q^2 \rangle \:\:(10^{-7}\:{\rm GeV^{-2}})$
                                          && $\wm0.775(94)\nb$   && $\wm0.517(60)\nb$               \\
$\langle F_L \rangle$                     && $\wm0.398(26)\nb$   && $\wm0.365(21)\nb$               \\
$\langle S_3 \rangle$                     && $-0.166(16)\nb$     && $-0.233(12)\nb$                 \\
$\langle S_4 \rangle$                     && $\wm0.3039(51)$     && $\wm0.3164(38)$                 \\
$\langle P_4^\prime \rangle$              && $\wm0.6223(91)$     && $\wm0.6582(46)$                 \\
\hline\hline
 \end{tabular}
\caption{\label{tab:binnedresults}Binned theoretical results in the Standard Model, for the two $q^2$ ranges
specified in the header of the table (in ${\rm GeV}^2$). The uncertainties given here are the total
uncertainties, as explained in the main text.}
 
\end{table}

We find that our Standard-Model results for the differential branching fractions of both $B^0 \to K^{*0}\mu^+\mu^-$ and
$B_s^0 \to \phi\, \mu^+ \mu^-$ are about 30\% higher than the experimental data. Note that for $B_s^0 \to \phi\, \mu^+ \mu^-$,
a higher-than-observed differential branching fraction was also found using form factors from light-cone sum rules \cite{Ball:2004rg}
(see Fig.~3 of Ref.~\cite{Aaij:2013aln}) and from a relativistic quark model \cite{Faustov:2013pca}. 
In the high-$q^2$ region considered here, our results for the
observables $F_L$, $S_5$, $P_5^\prime$, and $A_{FB}$ are in agreement with experiment.
For the $B^0 \to K^{*0}\mu^+\mu^-$ observables $S_3$, $S_4$, and $P_4^\prime$, we see deviations between the LHCb data
and our results in bin 1, in agreement with Refs.~\cite{Altmannshofer:2013foa, Hambrock:2013zya}.

\begin{figure}
\includegraphics[width=0.8\linewidth]{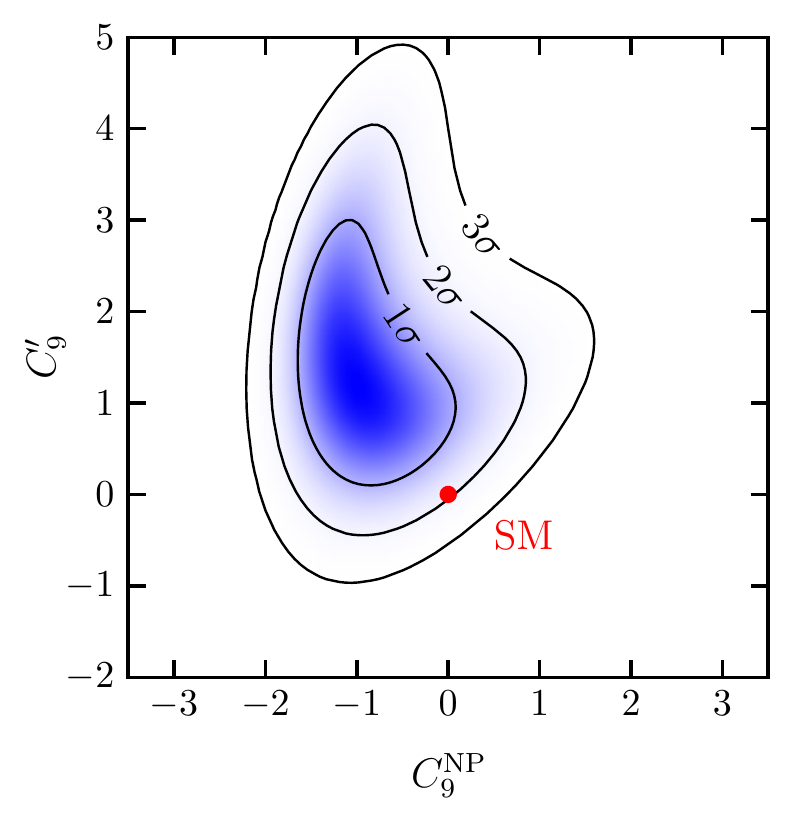}
\caption{\label{fig:fit}The likelihood function of a fit to the $B^0\to K^{*0}\mu^+\mu^-$ and $B_s^0\to \phi\,\mu^+\mu^-$
experimental data above $q^2=14.18\:{\rm GeV}^2$, with fit parameters $C_9^{\rm NP}$ and $C_9^\prime$. The contours
correspond to $\Delta \chi^2=2.30,\:6.18,\:11.83$.}
\end{figure}

To study the possibility of new physics in the Wilson coefficients $C_9$ and $C_9^\prime$, we performed a fit of these
two parameters to the experimental data above $q^2=14.18\:{\rm GeV}^2$, keeping all other Wilson coefficients fixed at their Standard-Model values
(and assuming $C_9,C_9^\prime\in\mathbb{R}$). We included the observables $\mathrm{d}\mathcal{B}/\mathrm{d}q^2$, $F_L$,
$S_3$, $S_4$, $S_5$, $A_{FB}$ for $B^0 \to K^{*0}\mu^+\mu^-$, and $\mathrm{d}\mathcal{B}/\mathrm{d}q^2$, $F_L$, $S_3$
for $B_s^0 \to \phi\, \mu^+ \mu^-$. We fully took into account the correlations between our theoretical results for different
observables and different bins. The best-fit values are
\begin{equation}
 C_9^{\rm NP}=-1.0\pm 0.6, \hspace{4ex} C_9^\prime = 1.2 \pm 1.0, \label{eq:fitresult}
\end{equation}
and the likelihood function is plotted in Fig.~\ref{fig:fit}. The dashed curves in Fig.~\ref{fig:observables} show the
observables evaluated at the best-fit values. To investigate how much the uncertainties in Eq.~(\ref{eq:fitresult})
are influenced by the theoretical and experimental uncertainties, we performed new fits where we artificially eliminated or reduced different sources of uncertainty.
In particular, setting all form factor uncertainties to zero results in $C_9^{\rm NP}=-0.9 \pm 0.4, C_9^\prime = 0.7 \pm 0.5$, and raises
the statistical significance for nonzero $(C_9^{\rm NP}, C_9^\prime)$ from 2$\sigma$ to 3$\sigma$. Reducing instead the experimental
uncertainties can have a more dramatic effect, because some of the angular observables already have very small theory
uncertainties compared to the current experimental uncertainties.

\vspace{3ex}

Our result (\ref{eq:fitresult}) is in remarkable agreement with the
result (\ref{eq:AltmannshoferGF}) of the fit performed in Ref.~\cite{Altmannshofer:2013foa}, which did not include the
$B_s^0 \to \phi\, \mu^+ \mu^-$ data. Equation (\ref{eq:fitresult}) is also consistent with the value $C_9^{\rm NP}\sim-1.5$
obtained in Ref.~\cite{Descotes-Genon:2013wba}, and with the very recent Bayesian analysis of Ref.~\cite{Beaujean:2013soa}.
As expected \cite{Altmannshofer:2013foa, Hambrock:2013zya}, the new-physics scenario (\ref{eq:fitresult})
does not remove the tension seen in bin 1 for $S_4/P_4^\prime$. Nevertheless, the fit (\ref{eq:fitresult})
significantly improves the overall agreement with the data, reducing the total $\chi^2$ by 5.7 and giving
$\chi^2/{\rm d.o.f.}=0.96$. We also performed a fit of the experimental data for all observables in bin 2 only, which gives
\begin{equation}
 C_9^{\rm NP}=-0.9\pm 0.7, \hspace{2ex} C_9^\prime = 0.4 \pm 0.7 \hspace{2ex}({\rm bin\:\:2\:\:only}). \label{eq:fitresultbin2}
\end{equation}

\newpage

A major concern about the calculations is the possibility of larger-than-expected contributions from broad
charmonium resonances above the $\psi(2S)$. In the $B^+ \to K^+ \mu^+ \mu^-$ differential decay rate, the LHCb Collaboration recently reported
sizable peaks associated with the $\psi(3770)$ and $\psi(4160)$ \cite{Aaij:2013pta}. Note that the OPE which we
use to include $c\bar{c}$ effects [Eq.~(\ref{eq:NLMEOPE})] is expected to describe only $q^2$-integrated observables (in the high-$q^2$ region) \cite{Beylich:2011aq}.
To test the robustness of our analysis,
we added Breit-Wigner amplitudes with the masses and widths of the $\psi(3770)$ and $\psi(4160)$ \cite{Ablikim:2007gd}
to $T_9(q^2)$, and included their complex-valued couplings as nuisance parameters. We constrained
the magnitudes of these couplings to allow the ratios of the purely resonant and nonresonant contributions
to the differential decay rates at $q^2=m_{\psi(3770)}^2$ and $q^2=m_{\psi(4160)}^2$ to be as large as
in Fig.~1 of Ref.~\cite{Aaij:2013pta}, but we left the phases unconstrained.
A fit of $C_9^{\rm NP}, C_9^\prime$ in the presence of these nuisance parameters gives
$C_9^{\rm NP}=-1.1 \pm 0.7, C_9^\prime = 1.2 \pm 1.1$; the significance for nonzero $(C_9^{\rm NP}, C_9^\prime)$
gets reduced to 1.4$\sigma$. We stress that adding Breit-Wigner amplitudes
is model-dependent and corresponds to a double counting of the $c\bar{c}$ degrees of freedom. A better
understanding of the resonant contributions from first-principles QCD is needed.

\textit{Acknowledgments:} We gratefully acknowledge discussions with Wolfgang Altmannshofer, William \mbox{Detmold},
Gudrun Hiller, Alexander Lenz, Joaquim Matias, Iain W.~Stewart, Jesse Thaler, and Michael Williams. SM is supported by the
U.S.~Department of Energy under cooperative research agreement Contract Number DE-FG02-94ER40818.
This work was supported in part by an STFC Special Programme Grant
(PP/E006957/1).  RH and MW are supported by an STFC Consolidated
Grant. ZL is partially supported by NSFC under the Project 11105153, the Youth Innovation Promotion Association
of CAS, and the Scientific Research Foundation for ROCS, SEM.

\end{document}